\documentclass[a4paper]{jpconf}
\usepackage{graphicx}
\begin{document}
\title{On the vibron-polaron damping in quasi 1D macromolecular chains}

\author{D. Cevizovic$^1$, A. Chizhov$^{2,3}$, A. Reshetnyak$^4$, Z. Ivic$^1$,
S. Galovic$^1$, \\ 
S. Petkovic$^1$}

\address{$^1$University of Belgrade, ``Vin\v ca'' Institute of Nuclear sciences,
Laboratory for Theoretical and Condensed Matter Physics, P.O. BOX 522, 11001,
Belgrade, Serbia\\
$^2$Joint Institute for Nuclear Research, Bogoliubov Laboratory of Theoretical
Physics, Dubna, 141980, Russia\\
$^3$Dubna State University, Dubna, 141980, Russia\\
$^4$Institute of Strength Physics and Materials Science SB RAS, Tomsk, 634055,
Russia}

\ead{cevizd@vinca.rs}

\begin{abstract}
The properties of the intramolecular vibrational excitation (vibron) in a quasi
1D macromolecular structure are studied. It is supposed that due to the vibron
interaction with optical phonon modes, a vibron might form partially dressed small
polaron states. The properties of these states are investigated in dependence
on the basic system parameters and temperature of a thermal bath. We also
investigate the process of damping of the polaron amplitude as a function of temperature and
vibron-phonon coupling strength. Two different regimes of the polaron damping
are found and discussed.
\end{abstract}

\section{Introduction}
The knowledge of exciton properties and especially physics of exciton
transport in quasi 1D structures is very important in many fields of
material sciences. Here, it is pertinent to mention the problem of charge and energy
transport in biological macromolecules (MCs) and the problem of the quantum
information transport in quantum circuits, especially in realistic conditions,
where a thermal bath is an important source of quantum information losses. For
example, the energy and charge transport in biological macromolecules
(such as protein macromolecules, RNA, and DNA) plays an important role in the
processes which take place in functioning of living cells. One of the most
important processes is the photosynthesis, in which a photon quantum is absorbed by
special cell structures (light-harvesting antennas) as an electronic excitation.
After that these excitations are transported to the reaction centres, where
they are converted to the chemical energy. It is remarkable fact that during its
migration almost every absorbed photon is successfully transferred to
the reaction centre despite its short life-time (the characteristic life-time
of an electron excitation is approximately 1 ns).

There are two main subsystems that we are interested in, namely a vibronic
excitation and phonons. Due to the presence of intersite coupling, a vibronic excitation
can migrate between two adjacent sites. The phonon subsystem
affects to the excitation migration on different ways. First, due to the presence
of the phonon-excitation interaction, the excitation can be self-trapped and
form a new quasiparticle dressed by the cloud of virtual phonons, i.e. a polaron
quasiparticle. Under some circumstances, such a quasiparticle can be more stable
than the bare excitation, and in principle such a model might explain the transfer of
the excitation over long distances along a macromolecular spine
\cite{DavydovJTB,DavydovPS,DavydovPD,AK}. Second, the phonon subsystem is a source
of dissipative processes in relation to the excitation (polaron) subsystem.
Phonons prevent the formation of the polaron quasiparticle and reduce its life-time.
As a consequence, the transfer properties of polarons in MCs are determined by the
interplay of dynamic and dissipative processes
\cite{DavydovPD,Petrov,Agranovich}.

In this paper, we investigate the process of relaxation of exciton states on the
lattice vibrations in 1D macromolecular chains. We suppose that, due to the
presence of exciton-phonon coupling, an exciton quasiparticle is self-trapped
and forms a non-adiabatic polaron state. It is then assumed that the
dissipation of such a state is the consequence of the presence of the residual
polaron-phonon interaction (all other sources of dissipation, such as
polaron-polaron interaction, are disregarded). The process of the exciton self-trapping
is described in the  Sec.~2. Main properties of the formed polaron state are presented
in dependence on the values of the basic energy parameters of macromolecular systems.
In the Sec.~3 we consider the dependence of the polaron amplitude damping on the system parameters
and the thermal bath temperature.

\section{Physical foundation of the problem}
The framework of our investigation is the Holstein molecular crystal model
(MCM)~\cite{Holstein}. We start from the Hamiltonian which includes the exciton (vibron)
interaction with the optical phonon subsystem.
\begin{equation}\label{SH}
H=E\sum_n{a^{\dagger}_na_n}-J\sum_n{a^{\dagger}\left(a_{n-1}+a_{n+1}\right)}
+\sum_q{\hbar\omega_qb^{\dagger}_qb_q}+\frac{1}{\sqrt{N}}\sum_{n,q}{
F_q\mathrm { e}^{iqnR_0}a^{\dagger}_na_n\left(b_q+b^{\dagger }_{-q}\right)} .
\end{equation}

In the above Hamiltonian $E$ is the exciton excitation energy, $J$ is the resonant transfer
integral (intersite coupling strength), which is responsible for quantum jumps
of the particle between neighbouring sites, $R_0$ is the parameter of the macromolecular
chain,  $a_n$ ($a^{\dagger}_n$) are the operators of annihilation (creation) of an
exciton on the $n$-th structure element of the macromolecular chain, and $b_q$
($b^{\dagger}_q$) are the phonon annihilation (creation) operators. The relation
for the exciton-phonon coupling constant $F_q$ depends on the nature of the phonon
subsystem. For the optical phonons, this parameter has the form
$F_q=\chi\sqrt{\hbar /(2M\omega_q)}$ with $\chi$ and $M$  being by  the interaction parameter
and the mass of the chain structure element respectively.

The standard procedure to pass into the polaron picture (dressed vibron point of view),
is based on the  Lang-Firsov (LF) unitary transformation \cite{AK,LF}. Such
approach is applicable in the case where the conditions for a formation of the
nonadiabatic polaron are fulfilled (i.e. for the strong-coupling
exciton--phonon case). However, in the case of a vibron excitation in the biological
macromolecules these conditions are not satisfied (and therefore vibron-phonon
interaction can belong to intermediate or weak coupling cases!)
\cite{PouthierJCP132}. Consequently, in this case, the variational approaches
are much more promising ones \cite{YarkonyCP,BI,CevizovicCPL2008,CevizovicPRE,
CevizovicCPB}. Here, we use the  variational approach based on the modified
Lang-Firsov unitary transformation (MLFUT):
$U=\exp (-\sum_{n} a^{\dagger}_na_nS_n)$, where
$S_n=1/\sqrt{N}\sum_q{f_q}\mathrm{e}^{-iqnR_0}(b_{-q}-b^{\dagger}_q)$,
and $f_q=f^*_{-q}$ being by the  parameters that determine the degree of exciton dressing
by a cloud of phonons. They can be obtained with use of the variational
approach, i.e. by the minimization of the polaron ground state energy. Using MLFUT,
we obtain a transformed Hamiltonian with new operators that correspond to
dressed exciton (polaron) operators and the new phonons (which describe
oscillations of structure elements around new equilibrium positions):
\begin{eqnarray}\label{TH1}
\tilde{H}&=&
\mathcal{E}\sum_n{a^{\dagger}_na_n}-J\sum_n{a^{\dagger}_n\left(a_{
n-1}\theta_{n-1}+a_{n+1}\theta_{n+1}\right)}+\sum_n{\hbar\omega_qb^{\dagger}
_qb_q}+\nonumber\\
&&
+\frac{1}{\sqrt{N}}\sum_{n,q}{(F_q-\hbar\omega_qf^*_q)\mathrm{e}^{
iqnR_0}a^{ \dagger} _na_n\left(b_q+b^{\dagger}_{-q}\right)}+\tilde{H}_R ,
\end{eqnarray}
where $\theta_{n\pm1}=\exp (S_{n\pm1}-S_n)$ is dressing
operator, and
$\mathcal{E}=E+N^{-1}\sum_q{\hbar\omega_q|f_q|^2}-N^{-1}\sum_q{F_q(f_q+f^*_{-q})}$
is the energy shift of dressed excitons. The fourth term in
(\ref{TH1}) describes the remaining interaction between the polaron and new phonon, 
which is responsible
for the phonon-induced polaron dissipation processes (i.e. for polaron amplitude
damping). The rest part of the Hamiltonian reads as
$\tilde{H}_R=N^{-1}\sum_{q,n,m\neq
n}{\left\{\hbar\omega_q|f_q|^2-F_q(f_q+f^*_q)\right\}\mathrm{e}^{iq(n-m)R_0}
a_na_ma^{\dagger}_na^{\dagger}_m}$ corresponds to the multi-polaron
processes.

In order to account for the influence of thermal fluctuations of structure
elements on polaron properties, we make averaging of our Hamiltonian over the new phonon 
subsystem to get
$\mathcal{H}=\mathcal{H}_0+\mathcal{H}_{rest}$.
Here, $\mathcal{H}_0=\mathcal{H}_{vib}+\tilde{H}_{ph}$ stands for the effective
mean-field Hamiltonian consists of effective vibron
($\mathcal{H}_{vib}=\left\langle \tilde{H}-\tilde{H}_{ph}\right\rangle_{ph}$)
and the phonon ($\tilde{H}_{ph}$) parts. Further,
$\mathcal{H}_{rest}=\tilde{H}-\tilde{H}_{ph}-\left\langle
\tilde{H}-\tilde{H}_{ph}\right\rangle_{ph}$ is the rest part of the interaction
between the dressed exciton and new phonon. The symbol $\left\langle\ \ \right\rangle_{ph}$
denotes the averaging over new-phonon ensemble. Due to the fact that relaxation
processes for the  phonons which forms the thermal bath are rather fast (with the typical time constants
in the  biological macromolecular chains of  order $10^{-14}\div 10^{-11}$ s) in
comparison with the characteristic time-scale of the transfer processes, they are
populated according to an equilibrium statistical distribution. The explicit form of
effective vibron Hamiltonian is
\begin{equation}\label{Hvibeffn}
\mathcal{H}_{vib}=\mathcal{E}\sum_n{a^{\dagger}_na_n}-J\mathrm{e}^{-W(T)}\sum_n{
\left\lbrace a^{\dagger}_na_{n-1}+a^{\dagger}_na_{n+1}\right\rbrace}
\end{equation}
and the rest part of the Hamiltonian has the form
\begin{eqnarray}\label{Hrest}
\mathcal{H}_{rest}&=&
-J\sum_n{\left\lbrace a^{\dagger}_na_{n-1}\left(\theta_{n-1}
-\left\langle\theta_{n-1}\right\rangle_{ph}
\right)+a^{\dagger}_na_{n+1}\left(\theta_{n+1}-\left\langle\theta_{n+1}
\right\rangle_{ph}\right)\right\rbrace}+\nonumber\\
&&
+\frac{1}{\sqrt{N}}\sum_{n,q}{\left(F_q-\hbar\omega_qf^*_q\right)\mathrm{e}^{
iqnR_0}a^{\dagger}_na_n(b_q+b^{\dagger}_{-q})},
\end{eqnarray}
where
$\left\langle\theta_{n\pm1}\right\rangle_{ph}=\mathrm{e}^{-W(T)}$,
$W(T)=N^{-1}\sum_q{\left|f_q\right|^2(2\nu_q+1)(1-\cos(qR_0))}$, is
dressing fraction and
$\nu_q=\left(\mathrm{e}^{\hbar\omega_q/k_BT}-1\right)^{-1}$ describes the  equilibrium
population of the $q$-th phonon mode. The first term in (\ref{Hrest})
represents the fluctuations of the vibron-phonon interaction
energy around its mean value. We assume that these fluctuations are small, and
they can be disregarded. The second part is the main part of the polaron-phonon
interaction which is responsible for the attenuation of the polaron amplitude.

We then transform our Hamiltonian from the coordinate to the wave vector $k$ representation
using $a_n=1/\sqrt{N}\sum_k{\mathrm{e}^{iknR_0}a_k}$, and
$a^{\dagger}_n=1/\sqrt{N}\sum_k{\mathrm{e}^{-iknR_0}a^{\dagger}_k}$ can be given as
\begin{equation}\label{Heffk}
\mathcal{H}=\mathcal{H}_{vib}+\tilde{H}_{ph}+\mathcal{H}_{rest}=\sum_k{\mathcal{
E}_ka^{\dagger}_ka_k}+\sum_q{\hbar\omega_qb^{\dagger}_qb_q}+\mathcal{H}_{rest},
\end{equation}
where
\begin{equation}\label{Hrestk}
\mathcal{H}_{rest}=\frac{1}{\sqrt{N}}\sum_{k,q}{
\left(F_q-\hbar\omega_qf^*_q\right)a^{\dagger}_{k+q}a_k(b_q+b^{\dagger}_{-q})} ,
\end{equation}
and
\begin{equation}
\mathcal{E}_k=\frac{1}{N}\sum_q{\hbar\omega_q|f_q|^2}-\frac{1}{N}\sum_q{
F_q(f_q+f^*_{-q})}-2J\mathrm{e}^{-W(T)}\cos(kR_0) \nonumber
\end{equation}
is the polaron energy band measured from the vibron excitation energy. At
this place, we will mention that the nature of self-trapped exciton state can
be characterized by the mutual ratio of three basic energy parameters: vibron
energy band width $2J$, characteristic phonon energy $\hbar\omega_C$, and
polaron binding energy $E_b=N^{-1}\sum_q |F_q|^2/(\hbar\omega_q)$.
In the process of vibron self-trapping in biological macromolecules, optical
phonon modes are of particular significance \cite{AK,BarthesJML,ScottPR217}.
For that reason, we restricted our calculations to the case of exciton
interaction with non-dispersive optical phonon modes
($\omega_q=\omega_0=\omega_C$). Standard small-polaron theory corresponds to
such systems where $E_b\gg\hbar\omega_C\gg 2J$, i.e. to the systems with narrow
energy band. In such structures, quasiparticle and phonon cloud forms new
entity, dressed quasiparticle, and formed deformation instantaneously follows
the motion of quasiparticle.

Further, in order to facilitate the analytic treatment, we assume that all
phonon modes equally influence to exciton dressing. In this case, we introduce
only one variational parameter $\delta$ ($0<\delta<1$), which can be defined as
$f_q=\delta\cdot F^*_q/(\hbar\omega_q)$. This parameter measures the degree of
the narrowing of the polaron band. Its value determines whether the vibron
motion takes place in the coherent fashion (when $\delta$ takes low values)
by means of the band mechanism, or it is achieved incoherently by the random
jumps between neighbouring sites (for high values of $\delta$). Although it is
technically less demanding, such an approach provides the results that are quite
satisfying, especially for system parameter values that are of our interests
\cite{CevizovicJPcs2012}.

Finally, the calculations are performed in terms of two system
parameters, namely the adiabatic parameter ($B=2J/(\hbar\omega_C)$) and
the coupling constant ($S=E_b/(\hbar\omega_C)$). In this case, we have
$W(\tau)=\delta^2S\coth{(1/2\tau)}$, where $\tau=k_BT/\hbar\omega_0$ is the
normalized temperature. Under these assumptions, the ground state polaron energy
($\mathcal{E}_{k=0}$) normalized to the characteristic phonon energy
$\hbar\omega_0$
takes the form
\begin{equation}\label{Egs}
\mathcal{E}_{gs}=S\delta(\delta-2)-B\mathrm{e}^{-S\delta^2\coth{(1/2\tau)}}.
\end{equation}

\begin{figure}[h]
\begin{center}
	\includegraphics[height=5 cm]{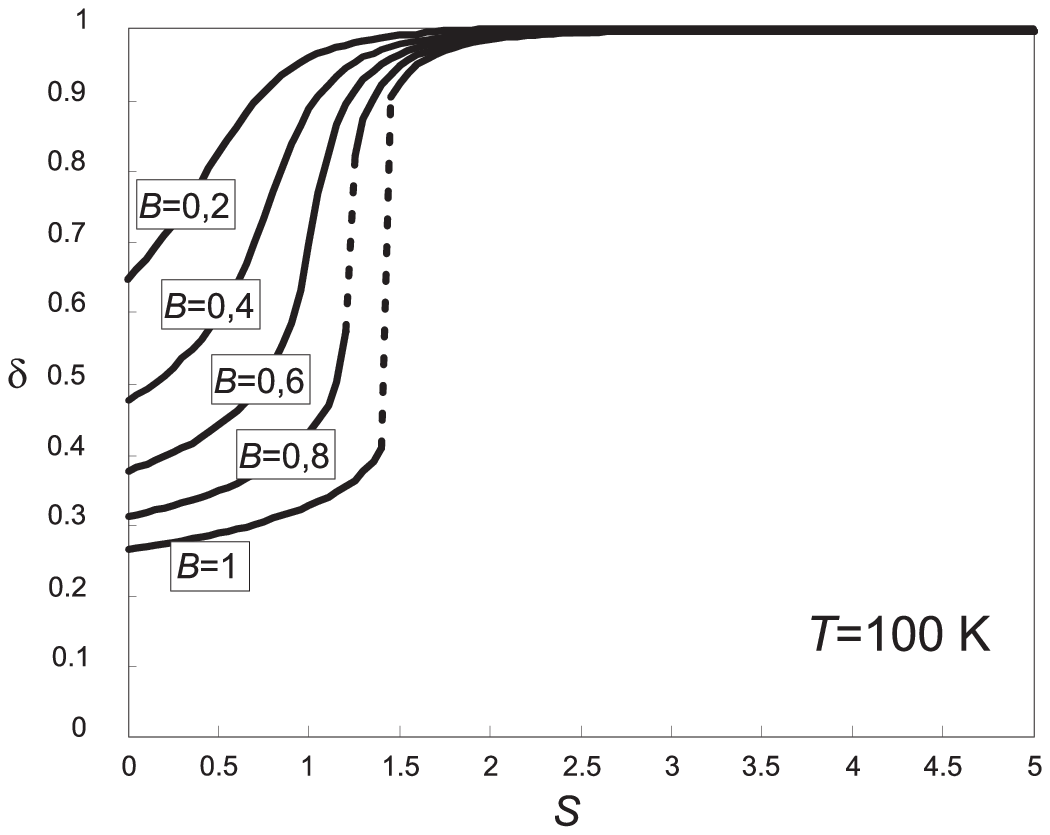}
	\hspace*{1cm}
	\includegraphics[height=5 cm]{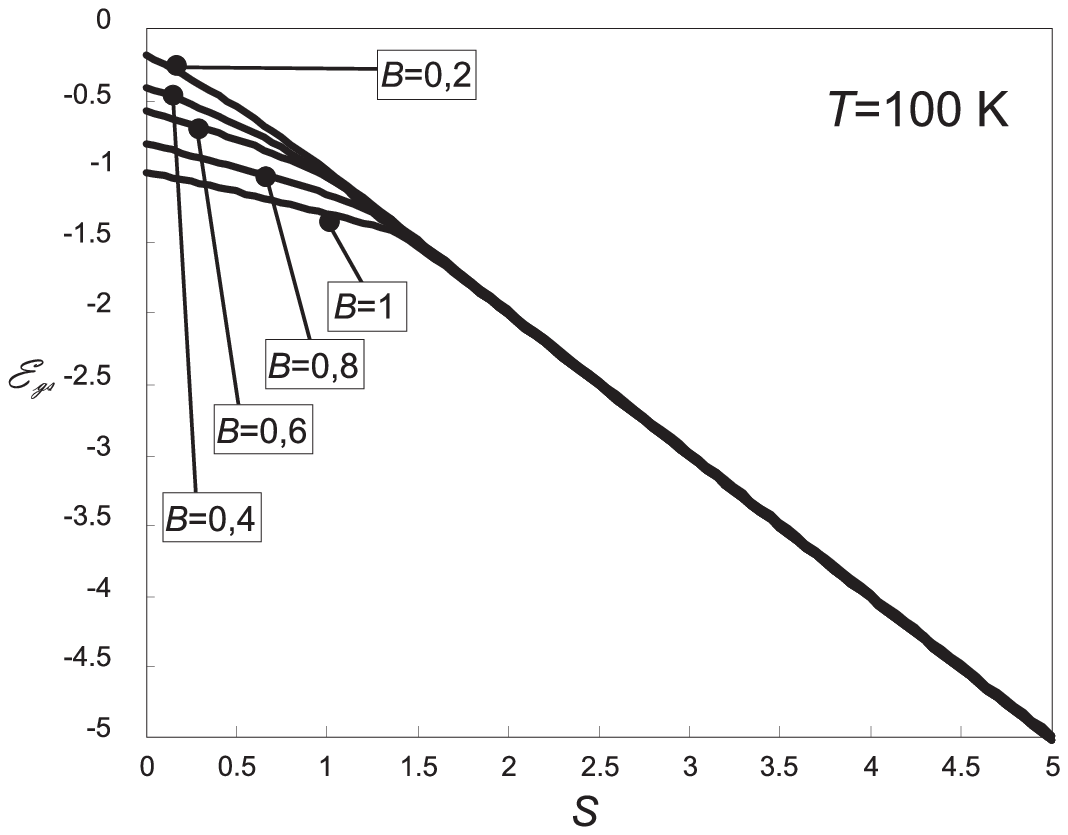}
  \caption{
\label{fig1}  
The dependence of the vibron dressing (left) and the ground state energy
(right) on the coupling constant for various vales of the adiabatic parameter at
temperature $T=100$~K. Critical values of the adiabatic parameter $B_C=0.68$ and
the coupling constant $S_C=1.3$ belong to moderately non-adiabatic and strong
coupling limits.}
\end{center}
\end{figure}

\begin{figure}[h]
\begin{center}
	\includegraphics[height=5 cm]{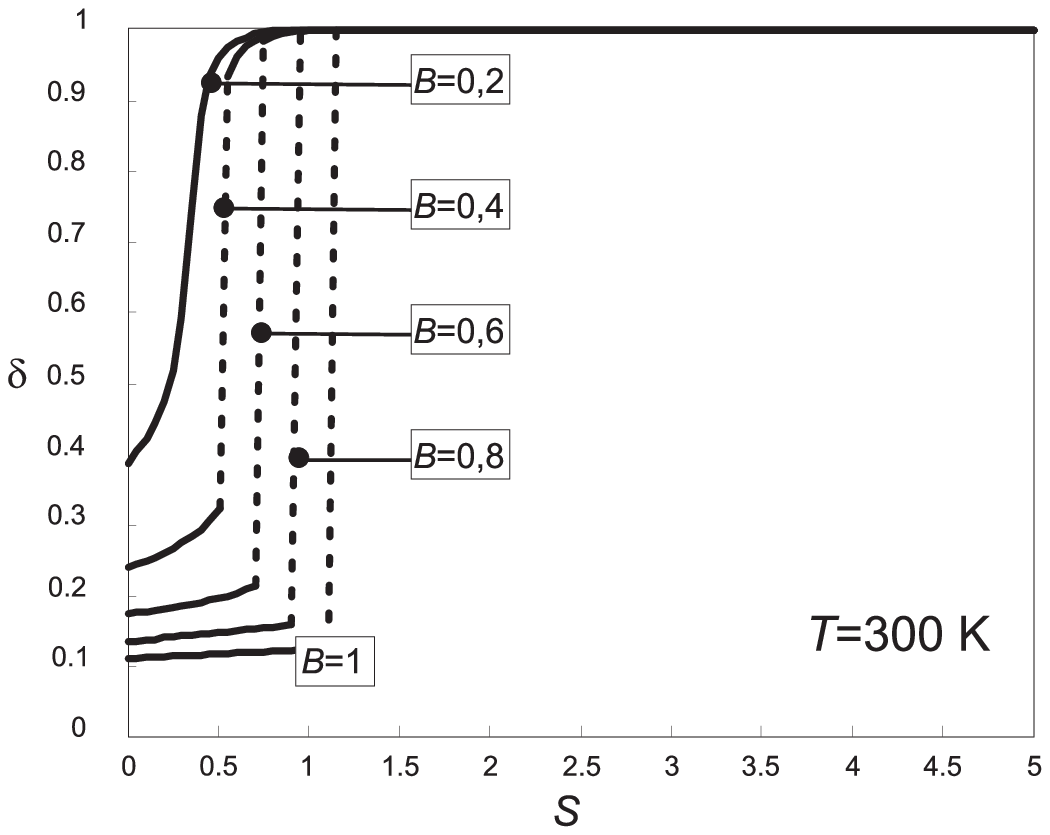}
	\hspace*{1cm}
	\includegraphics[height=5 cm]{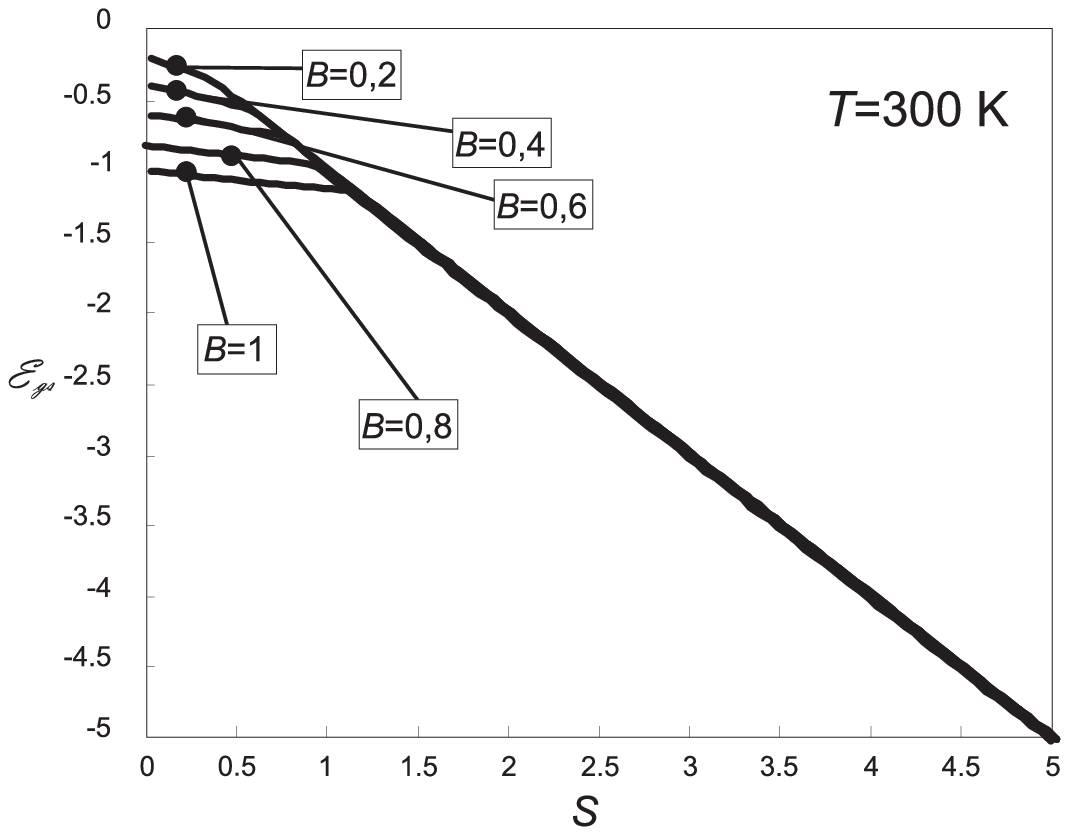}
  \caption{
\label{fig2}  
The dependence of the vibron dressing (left) and the ground state energy
(right) on the coupling constant for various vales of the adiabatic parameter at
temperature $T=300$~K. Critical values of the adiabatic parameter $B_C=0.22$ and
the coupling constant $S_C=0.4$ belong to non-adiabatic and weak coupling limits.}
\end{center}
\end{figure}

An optimal polaron state should be determined by the minimization
of the polaron ground state energy, from the condition
$\partial\mathcal{E}_{gs}/\partial \delta=0$. Consequently, the variational
parameter can be found by solving of the following transcendental equation
\begin{equation}\label{deltaOP}
\delta-1+\delta B\coth{(1/2\tau)}\mathrm{e}^{-S\delta^2\coth{(1/2\tau)}}=0.
\end{equation}
The dependence of the exciton dressing and its ground-state energy upon the basic
system parameters and temperature are presented in Figs.~\ref{fig1} and \ref{fig2}.

Obtained results appear to be consistent with the predictions of our earlier work
concerned with the vibron self-trapping in the alpha-helix and two parallel chain
structures \cite{CevizovicCPB,CevizovicPB2016}. Briefly, we found that the
system parameter space is divided on two characteristic areas: in the first area
vibron dressing is a continuous function of the coupling constant $S$ for fixed
values of $T$ and $B$. Such a behavior corresponds to the standard
small-polaron quasiparticle. In the second area the vibron excitation abruptly changes
its dressing:  being initially a slightly dressed (practically free)
quasiparticle, after an abrupt change its nature corresponds to the strongly dressed, 
practically immobile quasiparticle.
The border between these two areas (determined by the critical values $B_C$
and $S_C$) depends on the temperature of the bath: with increasing of the
temperature, its moves towards higher values of $S$ and $B$. It
means that for the low temperatures the so called critical values belong to non-adiabatic
($B\ll 1$) and weak ($S\ll 1$) limits. However, for the room temperatures, the critical values
may still belong to the non-adiabatic and intermediate-coupling! This result is
interesting for understanding of the nature of the vibron dressing mechanism which
is usually modelled in the framework of the adiabatic (soliton) polaron picture
\cite{DavydovJTB,DavydovPS,DavydovPD,BrizhikPRE}. On the other hand, it is
consistent with some numerical investigations provided by Hamm and Tsironis~\cite{HammTsironis}.

\section{Finding of the polaron amplitude damping}
Since the characteristic time-scale of the exciton transfer processes is much
larger than the time of phonon relaxation, in order to describe exciton
migration processes it is necessary to use the non-equilibrium statistical
distribution for exciton subsystem, i.e. the density matrix of excitons must be
described by non-equilibrium density operator. On the other hand, as it was
mentioned, the phonon subsystem is regarded as a subsystem in the equilibrium state and
it is described by the equilibrium statistical operator. Consequently, in order to
examine the polaron damping, we follow the well-known approach developed within the
theory of the open quantum systems \cite{ValasekTMF,Sadovskii}, and we derive
a set of evolution equations for non-equilibrium statistical average values
$\left\langle a_k\right\rangle=\mathrm{Tr}\left\langle\rho(t)a_k\right\rangle$
of polaron operators $a_k$. Here $\rho(t)=\rho_e(t)\otimes\rho_{ph}$ is the time
dependent non-equilibrium statistical operator (whereas  $\rho_e(t)$ is the non-equilibrium
time-dependent exciton statistical operator, while $\rho_{ph}$ is phonon
statistical operator). Evolution of the mean value of operator $A$ is determined by the
relation $i\hbar d\langle A \rangle /dt =\left\langle[A,H]\right\rangle$. For the Hamiltonian
(\ref{Heffk}) we obtain
\begin{equation}\label{dak}
i\hbar\frac{d}{dt}\left\langle a_k\right\rangle=\mathcal{E}_k\left\langle
a_k\right\rangle+\sum_q{\phi_q\left\lbrace\left\langle
a_{k-q}b_q\right\rangle+\left\langle
a_{k-q}b^{\dagger}_{-q}\right\rangle\right\rbrace},
\end{equation}
where $\phi_q=(F_q-\hbar\omega_qf^*_q)/\sqrt{N}$ 
(in the case of $\delta$ variational approach this parameter becomes
$\phi_q=F_q(1-\delta)/\sqrt{N}$). In the equation above, there appear two new amplitudes, namely 
$\left\langle a_{k-q}b_q\right\rangle$ and $\left\langle
a_{k-q}b^{\dagger}_{-q}\right\rangle$. The equation of motion for these amplitudes
contain other new amplitudes: $\left\langle a_kb_{q_1}b_{q_2}\right\rangle$,
$\left\langle a_kb^{\dagger}_{q_1}b_{q_2}\right\rangle$, $\left\langle
a_{k_1}a^{\dagger}_{k_2}a_{k_3}\right\rangle$, $\left\langle
a_kb_{q_1}b^{\dagger}_{q_2}\right\rangle$, and $\left\langle
a_kb^{\dagger}_{q_1}b^{\dagger}_{q_2}\right\rangle$. The equation of motion for
these amplitudes contain amplitudes that correspond to multiparticle
processes which probability is lower in order of magnitude than the probability
of single-particle processes. For this reason, it is possible
to cut-off the above mentioned amplitudes as follows: $\left\langle
a_kb_{q_1}b_{q_2}\right\rangle=\left\langle
a_k\right\rangle\left\langle b_{q_1}b_{q_2}\right\rangle=0,$ $\left\langle
a_kb^{\dagger}_{q_1}b^{\dagger}_{q_2}\right\rangle=\left\langle
a_k\right\rangle\left\langle
b^{\dagger}_{q_1}b^{\dagger}_{q_2}\right\rangle=0,$ $\left\langle
a_kb^{\dagger}_{q_1}b_{q_2}\right\rangle=\left\langle
a_k\right\rangle\left\langle b^{\dagger}_{q_1}b_{q_2}\right\rangle=\left\langle
a_k\right\rangle\nu_{q_1}\delta_{q_1,q_2},$ $\left\langle
a_kb_{q_1}b^{\dagger}_{q_2}\right\rangle=\left\langle
a_k\right\rangle\left\langle b_{q_1}b^{\dagger}_{q_2}\right\rangle=\left\langle
a_k\right\rangle (1+\nu_{q_1})\delta_{q_1,q_2}.$ One should have in mind that the
phonon subsystem is in the equilibrium state. Finally, we neglect all
the multipolaron processes $\left\langle
a_{k_1}a^{\dagger}_{k_1}a_{k_2}\right\rangle\approx 0$. Under these
assumptions the equations of motion for $\left\langle a_{k-q}b_q\right\rangle$,
and $\left\langle a_{k-q}b^{\dagger}_{-q}\right\rangle$ take the form:
\begin{equation}\label{dabk11}
i\hbar\frac{d}{dt}\left\langle
a_{k-q}b_q\right\rangle=(\mathcal{E}_{k-q}+\hbar\omega_q)\left\langle
a_{k-q}b_q\right\rangle+\phi_{-q}(\nu_q+1)\left\langle a_k\right\rangle ,
\end{equation}

\begin{equation}\label{dabk22}
i\hbar\frac{d}{dt}\left\langle
a_{k-q}b^{\dagger}_{-q}\right\rangle=(\mathcal{E}_{k-q}
-\hbar\omega_q)\left\langle
a_{k-q}b^{\dagger}_{-q}\right\rangle+\phi_{-q}\nu_q\left\langle a_k\right\rangle .
\end{equation}

After integration of Eqs.~(\ref{dabk11}) and (\ref{dabk22}), one can obtain
\begin{equation}\label{abk1}
\left\langle a_{k-q}b_q\right\rangle=-\left\langle
a_k\right\rangle\left\lbrace\mathcal{P}\left(
\frac{1}{\mathcal{E}_{k-q}+\hbar\omega_q}\right)+i\pi\delta(\mathcal{E}_{k-q}
+\hbar\omega_q)\right\rbrace\phi_{-q}(\nu_q+1) ,
\end{equation}

\begin{equation}\label{abk2}
\left\langle a_{k-q}b^{\dagger}_{-q}\right\rangle=-\left\langle
a_k\right\rangle\left\lbrace
\mathcal{P}\left(\frac{1}{\mathcal{E}_{k-q}-\hbar\omega_q}
\right)-i\pi\delta(\mathcal{E}_{k-q}-\hbar\omega_q)\right\rbrace\phi_{-q}\nu_q,
\end{equation}
where $\mathcal{P}(1/[\mathcal{E}_k\pm\hbar\omega_q])$ are the
principal values of corresponding integrals. After substitution of
Eqs.~(\ref{abk1}), (\ref{abk2}) into Eq.~(\ref{dak}), we obtain a Schr\"{o}dinger-like
equation with damping
\begin{equation}\label{dakf}
i\hbar\frac{d}{dt}\left\langle
a_k\right\rangle=(\mathcal{E}_k-\Delta_k)\left\langle
a_k\right\rangle-i\gamma_k\left\langle a_k\right\rangle
\end{equation}

Such an equation describes the non-equilibrium dynamics of polaron, and it determines
the polaron energy shift and polaron amplitude damping caused by the polaron
interaction with the environment (i.e. with the thermal oscillations of the
macromolecular chain). Here
\begin{equation}
\Delta_k=\sum_q{|\phi_q|^2\nu_q\mathcal{P}\left(\frac{1}{\mathcal{E}_{k+q}
-\hbar\omega_q
}\right)}+\sum_q{|\phi_q|^2(1+\nu_q)\mathcal{P}\left(\frac{1}{\mathcal{E}_{k+q}
+\hbar\omega_q}\right)}
\end{equation}
is the shift of the polaron energy, and
\begin{equation}
\gamma_k=\pi\left\lbrace{\sum_q{|\phi_q|^2\nu_q\delta\left(\mathcal{E}
_{k+q}-\hbar\omega_q\right)}+\sum_q{|\phi_q|^2(1+\nu_q)\delta\left(\mathcal{E}_{
k+q} +\hbar\omega_q\right)}}\right\rbrace
\end{equation}
is the damping factor of the polaron amplitude which determines the
characteristic relaxation time $\tau_k\sim 1/\gamma_k$. As can be remarked,
the polaron amplitude damping is caused by the emission and absorption of real
phonons, because the presence of the delta-functions in $\gamma_k$ shows that
the energy is conserved during such processes. On the other hand, the shift of
the polaron energy $\Delta_k$ arises as a consequence of the emission and
absorption of virtual phonons, because it is meaningful only in the case when
$\mathcal{E}_{k+q}\neq\pm\hbar\omega_q$.

\begin{figure}[h]
\begin{center}
	\includegraphics[height=5 cm]{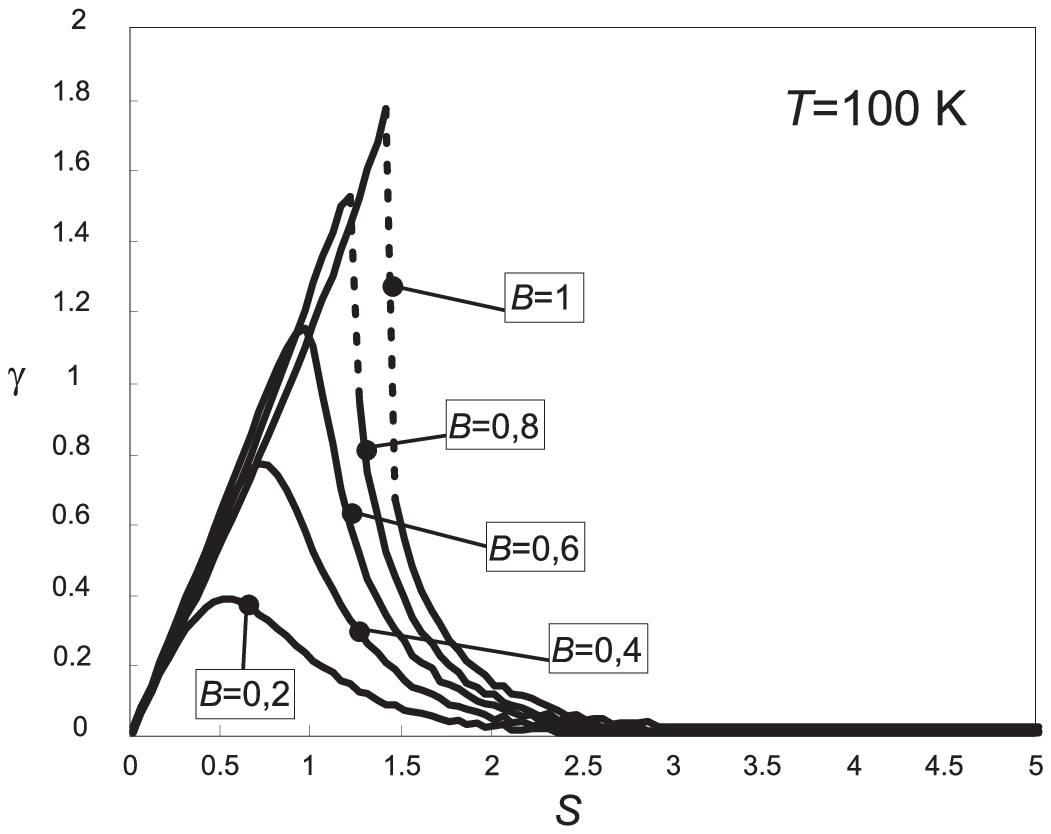}
	\hspace*{1cm}
	\includegraphics[height=5 cm]{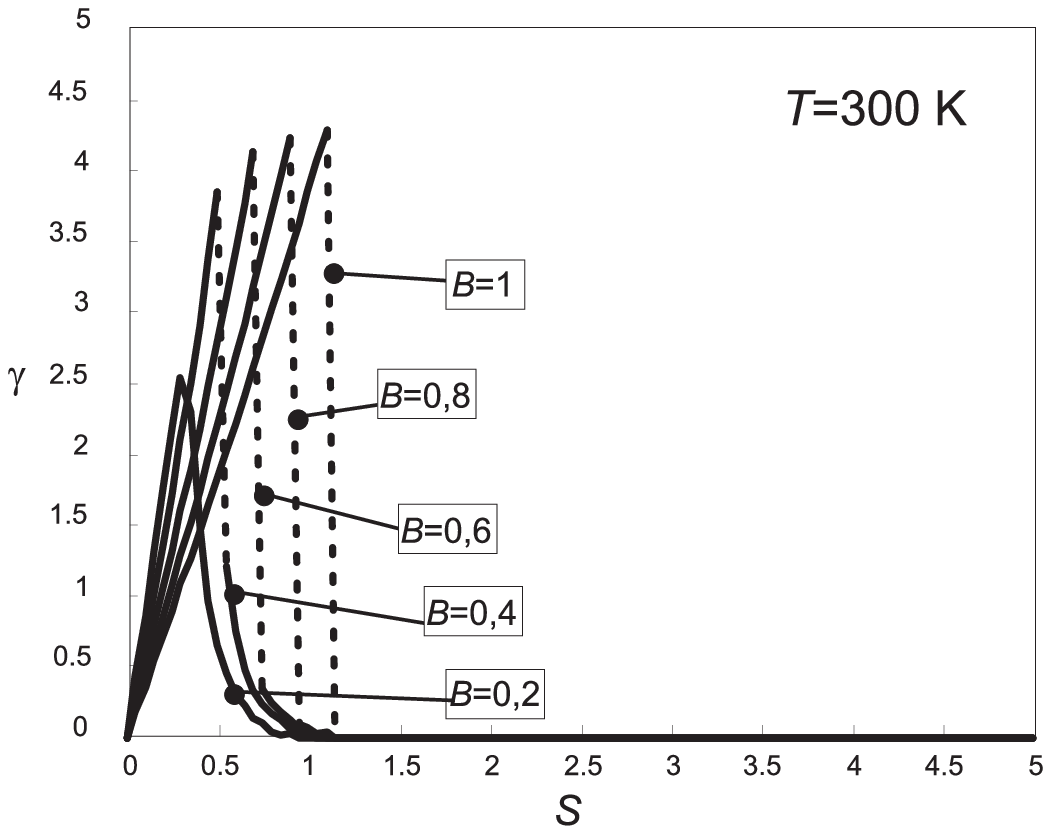}
  \caption{
\label{fig3}  
The dependence of the vibron damping on the coupling constant for
various vales of the adiabatic parameter at temperatures $T=100$~K (left) and
$T=300$~K (right).}
\end{center}
\end{figure}

In the case when the vibron interacts with non-dispersive optical phonon modes
the normalized energy of damped system looks as
\begin{equation}\label{Deltak}
\Delta_k=S(1-\delta)^2I_{\Delta} ,
\end{equation}
where integral $I_{\Delta}$ has the form
$$I_{\Delta}=\frac{1}{N}\sum_q{\frac{\coth(1/2\tau)\left[
S\delta(\delta-2)-B\mathrm{e}^{-S\delta^2\coth(1/2\tau)}\cos{(k-q)R_0}\right]-1}
{\left[S\delta(\delta-2)-B\mathrm{e}^{-S\delta^2\coth(1/2\tau)}\cos{(k-q)R_0}
\right]^2-1}},$$
and the damping factor becomes
\begin{equation}
\gamma_k=\pi E_b(1-\delta)^2I_{\gamma} ,
\end{equation}
where
$$I_{\gamma}=\frac{\mathrm{e}^{1/\tau}}{\mathrm{e}^{1/\tau}-1}\frac{1}{N}
\sum_q{\delta(\bar{\mathcal{E}}_{k-q}+1)}+\frac{1}{\mathrm{e}^{1/\tau}-1}\frac{1
}{N} \sum_q{\delta(\bar{\mathcal{E}}_{k-q}-1)}.$$
The dependence of the damping factor on the system parameters and temperature is shown
in Fig.~\ref{fig3}.

From the above results, one can remark that the damping of the vibron amplitudes induced by
the residual polaron-phonon interaction is significant in the area of the system
parameter space where low or partial vibron dressing takes place
(i.e., where $\delta < 1$). In the system parameter area where the exciton is slightly
dressed, the amplitude damping rises with the increasing of $S$. But in the area where
the exciton becomes heavy dressed, $\gamma$ sharply decreases. Such a behavior is
much more expressed for high $B$ (i.e., when the system becomes more adiabatic).
Additionally, this tendency increases with the rise of $B$, and for the system
parameters where the dressing abruptly changes, the damping becomes also discontinuous. 
The maximal value of $\gamma$ increases with increasing of $B$ and $T$.

Obtained results suggest that the increasing of system temperature can drive
the biological MCs to the system parameter area where the polaron crossover is
possible. In this area of the parameter space, the systems are quite sensitive to small
variations of temperature and the values of other system parameters. Amplitude
damping of slightly dressed vibrons (caused by residual vibron-phonon
interaction) in (moderately) non-adiabatic region strongly increases with the
increasing of the coupling constant and temperature of the thermal bath. In the nearly
adiabatic and adiabatic region such increasing is noticeably lower. At the same
time, the residual vibron-phonon interaction practically has no influence on
the amplitude damping of heavy dressed vibrons. The mean lifetime of nearly free
vibrons is very short (as compared to the mean lifetime in the low temperature limit), and
consequently the width of vibron spectral lines of the absorption spectra should
be high. These results could explain the absence of observation of weak
shift of the anomalous peak in the absorption spectra of crystalline acetanilide,
which was predicted by the use of the partially dressed SP picture.

\section*{Acknowledgments}
This work was supported by the Serbian Ministry of Education and Science under
Contract Nos. III-45010, III-45005 and by the Project within the Cooperation
Agreement between the JINR, Dubna, Russian Federation and Ministry of Education
and Science of Republic of Serbia.
One of the authors (A.Ch.) thanks the Blokhintsev-Votruba Program (JINR, Dubna)
for the financial support of participation in the conference ISQS25.

\end{document}